\documentclass{PoS}
\usepackage{amsmath} 
\usepackage{subcaption}
\usepackage{amssymb}

\title{$Z_b$ tetraquark channel and $B\bar B^*$  interaction  }

\ShortTitle{$Z_b$ tetraquark channel and $B\bar B^*$ interaction  }

\author{\speaker{Sasa Prelovsek}\\
       Faculty of Mathematics and Physics, University of Ljubljana,  Slovenia \\
        Jozef Stefan Institute, 1000 Ljubljana, Slovenia \\
        Instit\"ut f\"ur Theoretische Physik, Universit\"at Regensburg, D-93040 Regensburg, Germany \\
        E-mail: \email{sasa.prelovsek@ijs.si}}
        
        \author{H\"useyin Bahtiyar \\
        Jozef Stefan Institute, 1000 Ljubljana, Slovenia\\
        Department of Physics, Mimar Sinan Fine Arts University, Bomonti 34380, Istanbul, Turkey\\
        E-mail: \email{huseyin.bahtiyar@msgsu.edu.tr}}
        
        \author{Jan Petkovi\'  c  \\
          Department of Physics, University of Ljubljana, 1000 Ljubljana, Slovenia \\
        Jozef Stefan Institute, 1000 Ljubljana, Slovenia }

\abstract{  Two tetraquark candidates $Z_b(10610)$ and  $Z_b(10650)$ with flavor structure
 $\bar bb\bar du$ were discovered by Belle experiment in 2011.  We present  a preliminary $N_f=2$ lattice study of the $\bar bb\bar du$ system  in the approximation of static $b$ quarks, where the total spin of heavy quarks is fixed to one.   
 The ground and the excited eigen-energies    are determined as a function of separation $r$ between $b$ and $\bar b$.  The lower eigenstates  are related to a bottomonium and a pion. One of the higher eigenstates is  dominated by $B\bar B^*$: its energy is significantly below $m_B+m_{B*}$ for r=[0.1,0.4] fm, which suggests sizable attraction.  The  attractive potential $V(r)$ between $B$ and $\bar B^*$ is extracted assuming  that this eigenstate is related exclusively to $B\bar B^*$. Assuming  a certain form of the potential   and solving non-relativistic Schrodinger equation,   we find a   bound state pole  below $B\bar B^*$ threshold. For certain parametrizations, 
 the bound state is very close to the $B\bar B^*$ threshold   - this feature could be  related to $Z_b(10610)$ in the experiment.    }

\FullConference{37th International Symposium on Lattice Field Theory - Lattice2019\\
		16-22 June 2019\\
		Wuhan, China}

\begin{document}
 
 \section{Introduction}

The Belle experiment observed two  $Z_b^+$ states with exotic quark content $\bar bb\bar du$  and $J^P=1^+$ in 2011 \cite{Belle:2011aa,Garmash:2014dhx}.  The lighter state $Z_b(10610)$ lies slightly above $B\bar B^*$ threshold and the heavier $Z_b(10650)$ just above $B^*\bar B^*$ threshold. The experimentally discovered  decay modes are \cite{Belle:2011aa,Garmash:2014dhx,Garmash:2015rfd}
\begin{equation}
\label{channels}
Z_b^+ \to\ B\bar B^*,\ B^*\bar B^*, \Upsilon(1S)\pi^+,\  \Upsilon(2S)\pi^+,\ \Upsilon(3S)\pi^+,\ h_b(1S)\pi^+,\ h_b(2S)\pi^+
\end{equation}
 where the $B\bar B^*$ and  $B^*\bar B^*$ largely dominate   $Z_b(10610)$ and $Z_b(10650)$ decays, respectively.  
 
 The only preliminary lattice QCD study of these states has been reported in \cite{Peters:2016wjm,Peters:2017hon} and will be briefly reviewed below.   No other lattice results  are available since this channel present a severe challenge. The proper  method would require determination of scattering matrices for at least 7 coupled channels  (\ref{channels}) using the rigorous L\"uscher's method. Poles of such scattering matrix would render information on possible $Z_b$ states. Following this path seems too challenging for the moment. 
 
\begin{figure}[ht!]
\begin{subfigure}{0.24\textwidth}
\includegraphics[width=\linewidth]{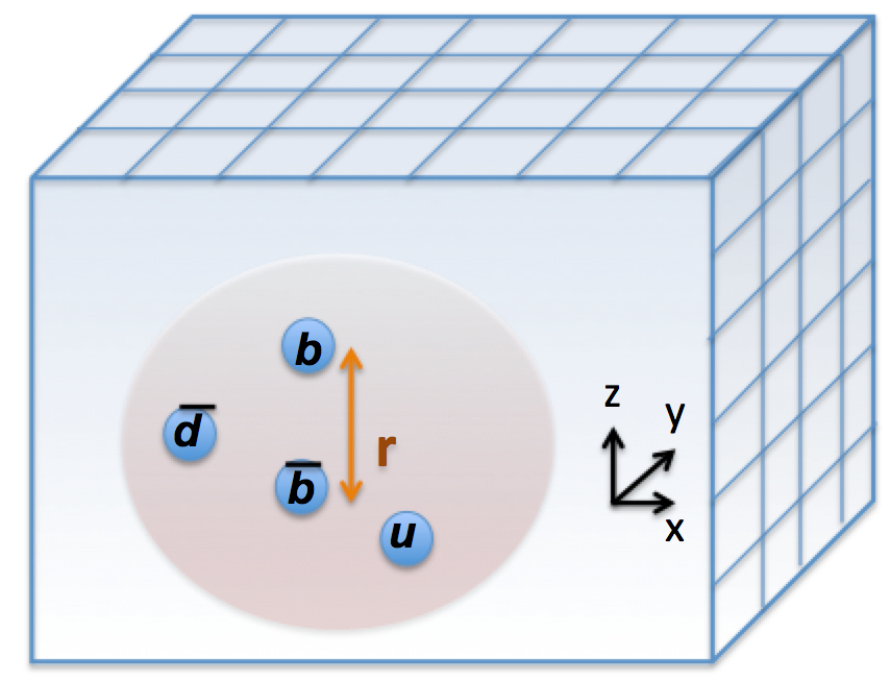}
\caption{ } \label{fig:1a}
\end{subfigure}
\hspace*{\fill} 
\begin{subfigure}{0.24\textwidth}
\includegraphics[width=\linewidth]{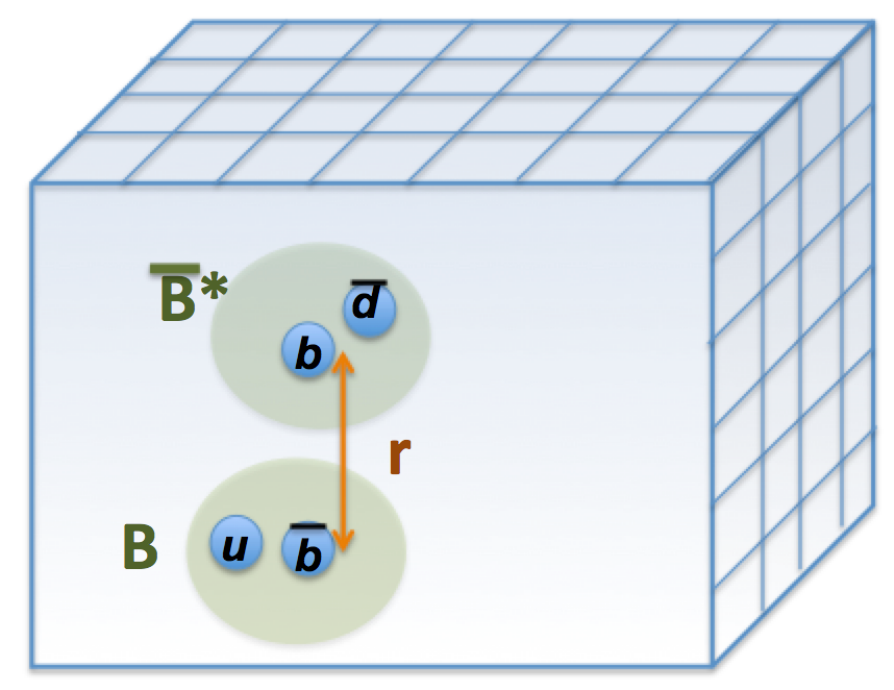}
\caption{ } \label{fig:1b}
\end{subfigure}
\hspace*{\fill} 
\begin{subfigure}{0.24\textwidth}
\includegraphics[width=\linewidth]{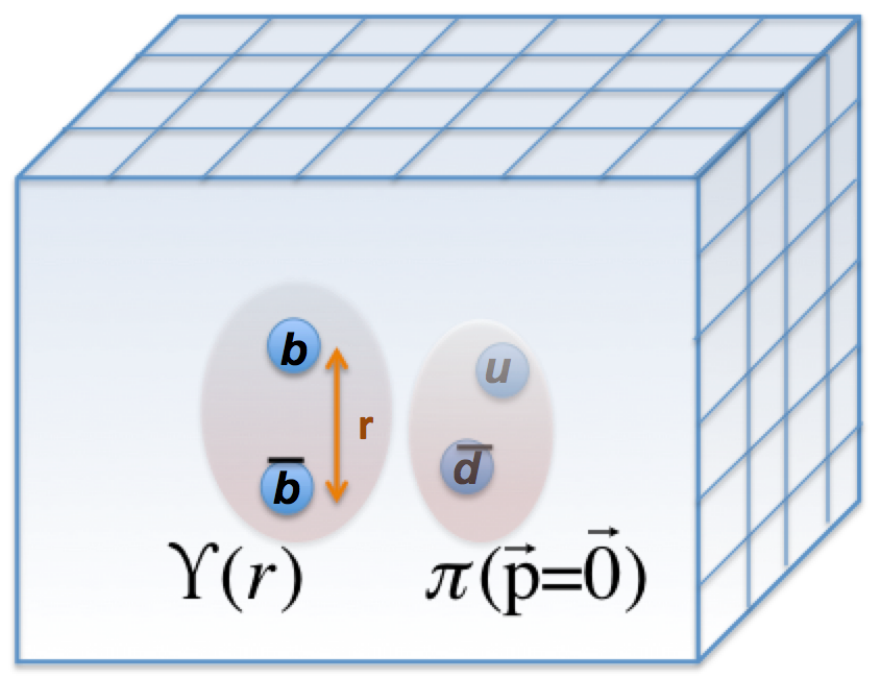}
\caption{ } \label{fig:1c}
\end{subfigure}
\begin{subfigure}{0.24\textwidth}
\includegraphics[width=\linewidth]{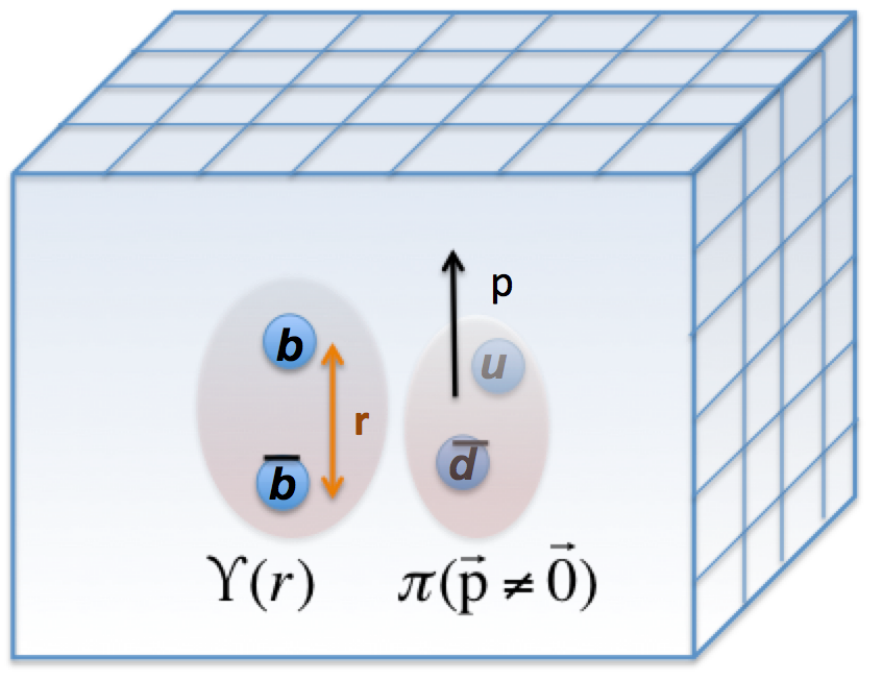}
\caption{ } \label{fig:1c}
\end{subfigure}
\caption{ (a) System considered. (b-d) Two-hadron Fock components relevant in the system with quantum numbers (\ref{sym}). } \label{fig:1}
\end{figure}

 In the present study we consider simplified Born-Oppenheimer approximation, inspired by the study of this system in \cite{Peters:2016wjm,Peters:2017hon}.  This approach is based on distinction between heavy and light degrees of freedom and finds enormous application 
 in molecular physics. It is valuable also for hadron systems with heavy quarks (see for example \cite{Braaten:2014qka,Brambilla:2017uyf}) and represents a good approximation for $Z_b$ system since $m_b$ is large.  
 In the first step,  $b$ and $\bar b$ are treated as static and fixed at distance $r$ (Figure \ref{fig:1a}).  We determine the  eigen-energies  $E_n(r)$ for the light degrees of freedom (light  $u/d$ quarks and gluons) in presence of these two static sources   as function of $r$ by lattice QCD.  The low-lying eigenstates (relevant for quantum numbers discussed in Section \ref{sec:qn}) are related to two-hadron states illustrated in Figs. \ref{fig:1} (b-d)
 \begin{equation}
 \label{two-hadron-states}
 B(0)\bar B^*(r),\ \  \Upsilon(r)~\pi(\vec p=0),\  \  \Upsilon(r)~\pi(\vec p\not =0),\  \  \Upsilon(r)~b_1(\vec 0)\ ,\qquad \vec p=\vec n \tfrac{2\pi}{L}~.
 \end{equation}
   The energy of $B\bar B^*$   in Fig.  \ref{fig:1}b   is of most interest  since $Z_b$ resonances lie near $B\bar B^*$ threshold.  This eigen-energy $E(r)$ provides the potential $V(r)=E(r)-m_B-m_{B*}$  between $B$ and $\bar B^*$ within certain simplifying assumptions mentioned below. The ground state of this system is not represented by $B\bar B^*$ (\footnote{In contrast to open beauty channel $bb\bar u\bar d$ where $BB^*$ is the lowest two-hadron channel. }), but by the   bottomonium-pion states $\Upsilon(r)\;\pi(\vec p)$. The $\Upsilon(r)$ denotes spin-one bottomonium where $\bar b$ and $b$ are separated by $r$ and its energy is given by the well-known static potential $V_{\bar bb}(r)$.  Pion can have zero or non-zero momentum  $\vec p=\vec n \tfrac{2\pi}{L}$ since total momenta of light degrees of freedom is not conserved in presence of static quarks, i.e. the momentum of light meson  is not conserved when it scatters on infinitely heavy bottomonium.  Our task is to extract energies of all these eigenstates and extract the potential $V(r)$ between $B$ and $\bar B^*$. 
   
   In the second step of the Born-Oppenheimer approach, the heavy degrees of freedom are relaxed from their static positions.  The $B$ and $\bar B^*$ mesons with finite masse $m_{B^{(*)}}^{exp}$  evolve in the extracted potential $V(r)$. The non-relativistic Schrodinger equation is solved to look for possible  $Z_b$ bound states, virtual bound states or resonances in this system. 
       
   The only previous lattice study of this system  \cite{Peters:2016wjm} presents preliminary results based on  Fock components  $B\bar B^*$ and $\Upsilon \pi(0)$. The presence of states  $\Upsilon \pi(\vec p\not =0)$  was mentioned in  \cite{Peters:2017hon}, but not included in the simulation. 
   
   \section{Quantum numbers and operators}\label{sec:qn}
   
   The quantum numbers of neutral $Z_b^0$ in experiment are $I=1$, $I_3=0$ and $J^{PC}=1^{+-}$.   Certain quantum numbers are somewhat different  in the systems with two static particles and are  reminiscent of diatomic molecule. We study the system  with quantum numbers  
      \begin{equation}
   \label{sym}
I=1, \ I_3=0,\quad S^{heavy}=1,\ S_z^{heavy}=0,\quad J_z^{light}=\lambda=0,\quad  \epsilon=-1,\quad  C\cdot P=-1~,
   \end{equation}
   which couple to $Z_b$ channel. 
   The neutral system is considered so that C-conjugation can be applied (Fig. \ref{fig:1} shows the charged partner).  Only the z-component of angular momenta for light degrees of freedom ($J_z^{light}$) is conserved. The $\epsilon$ is a quantum number related to the reflection over the yz plane. $P$ refers to inversion with respect to mid-point between $b$ and $\bar b$ and $C$ is charge conjugation, where only their product is conserved.  
   
   The spin  of infinitely heavy quark   can not flip via interaction with gluons, so $S^{heavy}$ of $\bar bb$  is conserved. We choose to study  system with $S^{heavy}=1,\ S_z^{heavy}=0$, where decays to $\Upsilon$ are allowed, while decays to $\eta_b$ and $h_b$ are forbidden. Note that  physical system $Z_b$ and $B\bar B^*$ with finite $m_b$ can be a linear combination of $S^{heavy}=1$ as well as $S^{heavy}=0$ and we  study only $S^{heavy}=1$ component. We have in mind this  component including $B\bar B^*$, $\bar BB^*$, $\bar B^* B^*$ ($O_1$ in Eq. \ref{operators}) when we refer to "$B\bar B^*$". 
   
   We employ 6 operators with quantum numbers (\ref{sym}) which resemble Fock components (\ref{two-hadron-states})  in Figs. \ref{fig:1} (b-d)
   \begin{align}
   \label{operators}
O_1= O^{B\bar B*}&\propto\sum_{a,b}\sum_{A,B,C,D}~\Gamma_{BA}\tilde \Gamma_{CD}~\bar b^a_C(0)q_A^a(0)~ \bar q^b_B(r)b_D^b(r)\ , \quad  \Gamma=P_-\gamma_5\ \tilde \Gamma=\gamma_z P_+,\\
&\propto [\bar b(0) P_- \gamma_5 q(0)]~[\bar q(r) \gamma_z P_+ b(r)]+[\bar b(0) P_- \gamma_z q(0)]~[\bar q(r) \gamma_5 P_+b(r)] \nonumber \\
O_2= O^{B\bar B*}&\propto\sum_{a,b}\sum_{A,B,C,D}~\Gamma_{BA}\tilde \Gamma_{CD}~\bar b^a_C(0)\nabla^2q_A^a(0)~ \bar q^b_B(r)\nabla^2b_D^b(r)\ , \quad  \Gamma=P_-\gamma_5\ \tilde \Gamma=\gamma_z P_+,\ \nonumber\\
O_3=O^{\Upsilon \pi(0)}&\propto\Upsilon_z~ \pi_{p=000} =[\bar b(0) \gamma_z P_+ b(r)]~[\bar q \gamma_5q]_{p=000}\nonumber\\
O_4=O^{\Upsilon \pi(1)}&\propto\Upsilon_z~ (\pi_{p=001}+\pi_{p=00-1}) =[\bar b(0) \gamma_z P_+ b(r)]~ \bigl([\bar q \gamma_5q]_{p=001}+ [\bar q \gamma_5q]_{p=00-1}\bigr)\nonumber\\
O_5=O^{\Upsilon \pi(2)}&\propto\Upsilon_z~ (\pi_{p=002}+\pi_{p=00-2}) =[\bar b(0) \gamma_z P_+ b(r)]~ \bigl([\bar q \gamma_5q]_{p=002}+ [\bar q \gamma_5q]_{p=00-2}\bigr)\nonumber\\
O_6=O^{\Upsilon ~b1(0)}&\propto\Upsilon_z~ (b1_z)_{p=000} =[\bar b(0) \gamma_z P_+ b(r)]~[\bar q \gamma_x\gamma_y q]_{p=000}\nonumber
\end{align}
and   color singlets are denoted by $[..]$.   First line in $O_{1,2}$  decouples spin indices of light and heavy quarks, so that transformation under $J^{light}_z$ is more transparent \cite{Peters:2016wjm} , while the second line in $O_1$ is obtained via Fierz transformation. $O_{4,5}$ have pion momenta  in $z$ direction due to $J^{light}_z$ and have two terms due to $C\cdot P$. The $\Upsilon\; b_1$ is not a decay mode  for finite  $m_b$ where $C$ and $P$ are separately conserved, but it is has quantum numbers (\ref{sym}) for $m_b\to \infty$.  All light quarks $q(x)$ are smeared around the central position $x$ using full distillation \cite{Peardon:2009gh} with radius $\simeq 0.3~$fm, while heavy quarks are point-like. 

We verified that there are no other two-hadron Fock components in addition  to (\ref{two-hadron-states}) with quantum numbers (\ref{sym}) and with non-interacting energies  below $m_B+m_{B^*}+0.2~$GeV.  

   \section{Lattice details}
   
 Lattice simulation with $N_f=2$, $m_\pi \simeq 266~$MeV and $a\simeq 0.124~$fm is performed. We take an ensemble with small $N_L=16$ and $L\simeq 2~$fm so that  only $\Upsilon \pi(p_z)$ with   $p_z\leq 2 \tfrac{2\pi}{L}$ appear in the energy region below $m_B+m_{B*}+0.2~$GeV; larger volumes would require additional operators like $O_{4,5}$ for higher pion momenta. 
 
Correlation matrices $\langle O_i (t)O_j^\dagger(0)\rangle$ for operators $O_{i,j=1,..,6}$ (\ref{operators}) are evaluated using full distillation method  \cite{Peardon:2009gh} and $\bar bb$ annihilation is omitted. The eigen-energies of the system are extracted from the correlation matrices using the well-known GEVP variational  approach.

    \begin{figure}[h!]
	\centering
	\includegraphics[width=0.6\textwidth]{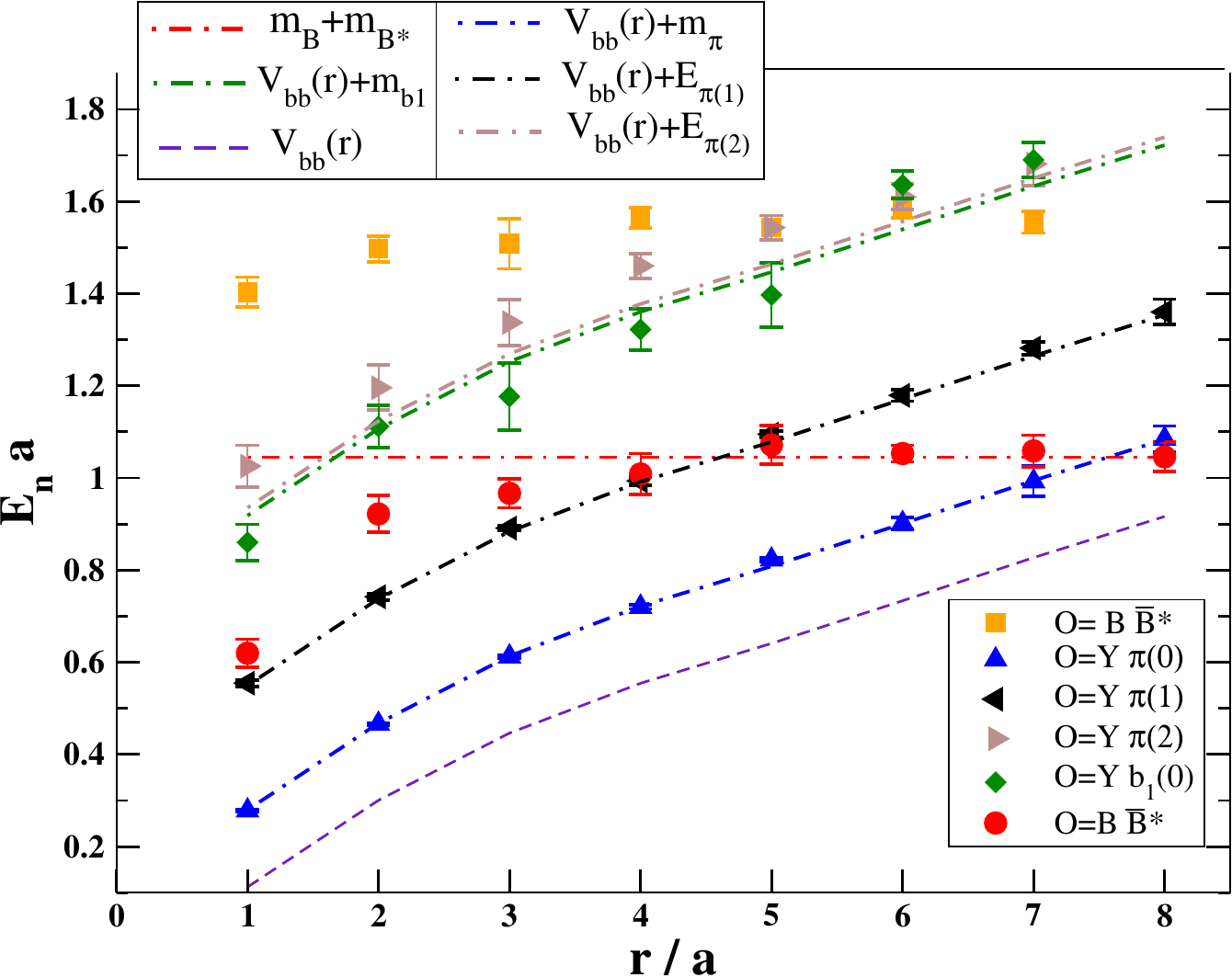}   
		\caption{ Eigen-energies of  $\bar bb\bar du$  system (Fig. 1a) for various separations $r$ between  static quarks $b$ and $\bar b$ are shown by points. The  label on the right indicates which two-hadron Fock component dominates each eigenstate. The dot-dashed lines indicate related two-hadron energies $E^{n.i.}$ (\ref{Eni}) in the limit when two hadrons  (\ref{two-hadron-states}) do not interact. The most important conclusion based on this spectra is that $B\bar B^*$ eigenstate (red crosses) has energy  significantly below $m_B+m_{B*}$, therefore shows  sizable strong attraction  for  $r=[0.1,0.4]~$fm. Lattice spacing is $a\simeq 0.124~$fm.  }
	\label{fig:En}
\end{figure}  

   \section{Eigen-energies of  $\bar bb\bar du$  system as a function of $r$}
   
   The main result of our study are the  eigen-energies of the $\bar bb\bar du$  system (Fig.  \ref{fig:1}a) with static $b$ and $\bar b$ separated by $r$, that are shown by points in Figure \ref{fig:En}.  The colors of points indicate which Fock-components among (\ref{two-hadron-states}) dominates an eigenstate, as determined from overlaps $\langle O_i|n\rangle$ of an eigenstate $|n\rangle$ to operator $O_i$. 
   
   The dashed lines provide related non-interacting (n.i.) energies $E_n$ of two-hadron states (\ref{two-hadron-states}) 
   \begin{equation}
   \label{Eni}
  E^{n.i.}_{B\bar B^*}=m_B+m_{B*},\quad   E^{n.i.}_{\Upsilon\pi(\vec p)}=V_{\bar bb}(r)+E_{\pi(\vec p)}=V_{\bar bb}(r)+\sqrt{m_\pi^2+\vec p^2}, \quad  E^{n.i.}_{\Upsilon b_1(0)}=V_{\bar bb}(r)+m_{b_1} 
   \end{equation} 
   where $\bar bb$ static potential $V_{\bar bb}(r)$, $m_B=m_{B^*}=0.5224(14)$ (for $m_b\to \infty$), $m_\pi$ and $m_{b_1}$ are determined on the same ensemble. 
       
      The eigenstate dominated by $B\bar B^*$ has energy close to $m_B+m_{B^*}$ for $r>0.5~$fm, but has it has significantly lower energy for $r\simeq [0.1,0.4]~$fm (red crosses in Fig. \ref{fig:En}). This indicates sizable strong attraction between $B$ and $\bar B^*$ in this system - something that might   be related to the existence of $Z_b$ tetraquarks. This is the most important and robust result of our lattice simulation. 

Other eigenstates are dominated by $\Upsilon \pi(\vec p)$ and $\Upsilon b_1$ states. Their  energies $E$  lie close to non-interacting energies $E^{n.i.}$ (\ref{Eni}) given by dot-dashed lines, so $E\simeq E^{n.i.}$. We point out that our simulation is not accurate enough to claim nonzero energy shifts $E-E^{n.i.}$ for $\Upsilon \pi$ and $\Upsilon b_1$ states, although Fig.   \ref{fig:En} shows small deviation 
from zero in some cases.

\section{Towards masses of $Z_b$ states within several simplifying approximations}\label{sec:V}

The total energy of the $\bar bb\bar du$ system is composed of the energy $E_n(r)$   for static $b$ and $\bar b$, determined in the previous section, plus the kinetic energy of heavy quarks, which presents a small perturbation within Born Oppenheimer approximation. The heavy quarks are now relaxed from their static positions and  evolve in the potentials determined from $E_n(r)$. 

\begin{figure}[tb!]
\begin{center} 
\includegraphics[width=0.6\textwidth]{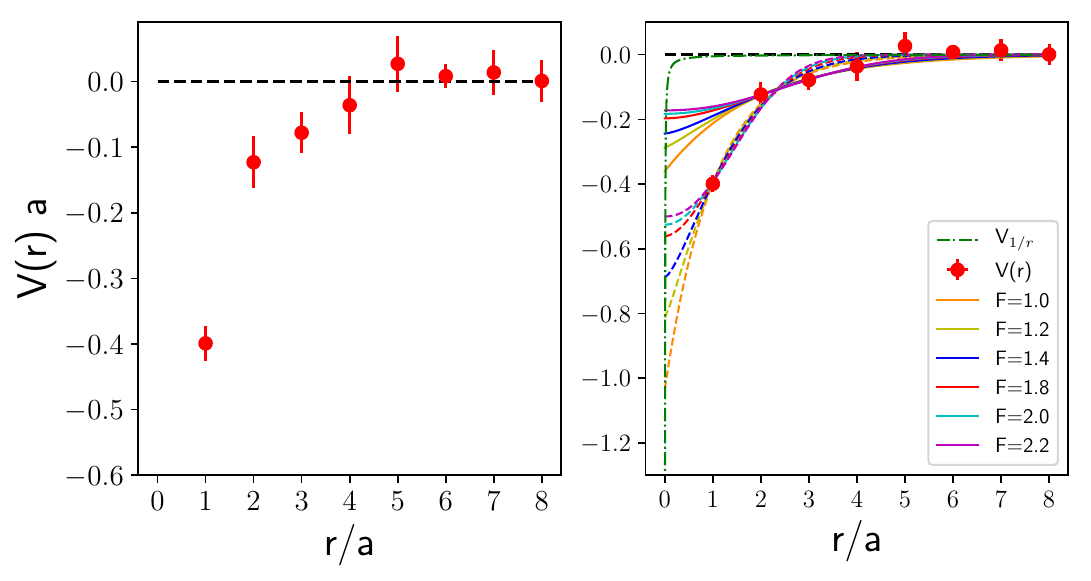}  
\caption{ (a) The   extracted potential $V(r)$ between $B$ and $\bar B^*$ from lattice. (b) Fits  of  $V(r)$ assuming the form   (\ref{V}) are presented by the solid and dashed lines  for various values of parameter $F$. Solid and dashed lines correspond to fits in the ranges $r/a=[2,4]$ and $[1,4]$, respectively.    
The singular potential $V_{1/r}(r)$ is shown by dot-dashed green line.  Lattice spacing is $a\simeq 0.124~$fm.   } \label{fig:V}
\end{center}
\end{figure}

We apply two serious simplifying approximations in order to shed some light on the possible existence of $Z_b$ based on energies  in Figure \ref{fig:En}. The first assumption is  that the eigenstate indicated by red crosses in Fig. \ref{fig:En} is related exclusively to $B\bar B^*$ Fock component and does not contain other Fock components in (\ref{two-hadron-states}). This is supported  by our lattice results
 to a good approximation, since this eigenstate couples almost exclusively to $O^{B\bar B^*}$  and has much smaller coupling to $O^{\Upsilon \pi}$ and $O^{\Upsilon b_1}$. In this case, the energy $E(r)$ of this eigenstate provides potential $V(r)=E(r)-m_B-m_{B*}$ between $B$ and $\bar B^*$, given in Fig. \ref{fig:V}. The potential shows sizable attraction for  small distances and is compatible with zero for $r\geq 0.6~$fm  within sizable statistical errors of our result. Lattice study that would probe whether one-pion exchange dominates at large $r$ would therefore need much higher accuracy. 
  
  The form of the potential $V(r)$ at $r<a\simeq 0.124~$fm is not known and it might be affected by discretization effects at $r\simeq a$. This brings us to the second simplifying approximation, where we assume a certain form to fit our potential in Fig. \ref{fig:V}
 \begin{equation}
 \label{V}
 V(r)=   -A\, e^{-(r/d)^F} .
 \end{equation}
 The fits of the lattice potential for   various choices of the parameter $F$  (\ref{V})  are shown in Fig. \ref{fig:V}. We employ two choices of fitting ranges   $r/a=[2,4]$ or $[1,4]$  since the  lattice potential at $r/a=1$ can be prone to the lattice discretization errors. 
 
   \begin{figure}[tb!]
\begin{center} 
\includegraphics[width=0.7\textwidth]{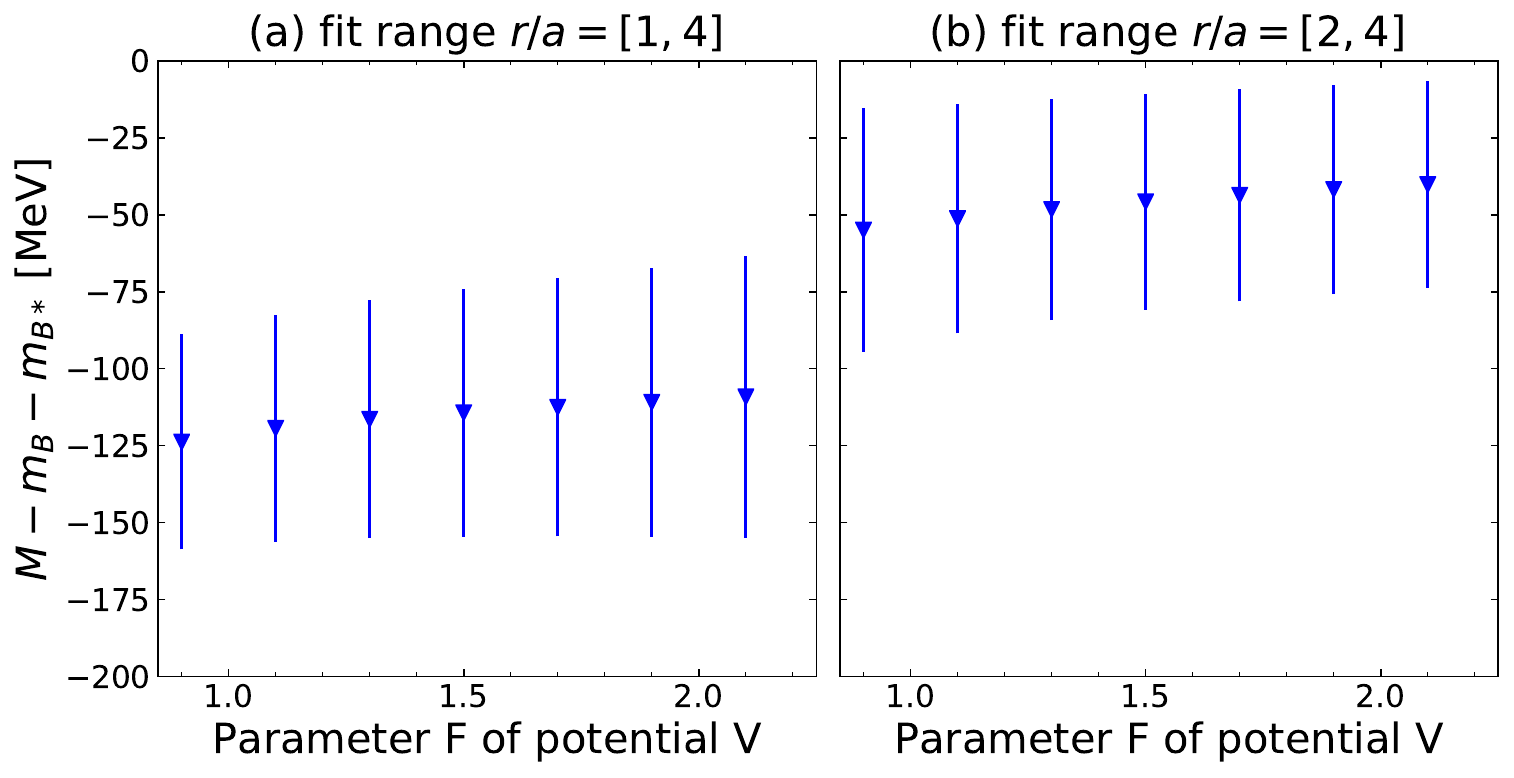}  
\caption{ The mass of the  bound state for various fits of the potential $V(r)$.   The mass is shown  for various choices of the parameter $F$ in $V(r)$ (\ref{V})  and the fitting range in $r$. } \label{fig:masses}
\end{center}
\end{figure}

The motion of $B$ and $\bar B^*$ within the extracted   potential $V(r)$ is analyzed by solving the non-relativistic 3D Schr\"odinger equation $[-\frac{1}{2\mu}\tfrac{d^2}{dr^2}+\tfrac{l(l+1)}{2\mu r^2}+V(r)]u(r)=Wu(r)$  for the experimentally measured  $B^{(*)}$ meson masses and $1/\mu=1/m_B^{exp}+1/m_{B^*}^{exp}$.    Here  $W=E^{tot}-m_B-m_{B^*}$ is the energy with respect to $B\bar B^*$ threshold.  The $B$ and $\bar B^*$ can couple to $Z_b$ channel  with $J^P\!=\!1^+$ in partial waves $l=0,2$.   We consider only $l=0$  since $V(r)+\tfrac{l(l+1)}{2\mu r^2}>0$   is repulsive for all $r$. We find one bound state   below threshold and its mass is shown   in Fig.    \ref{fig:masses}.  
The mass is presented for various choices of  the parameter $F$ in  the potential  (\ref{V}) and the fitting ranges in $r$. 
 The bound state with mass $M$ lies at  
\begin{equation}
\label{mass}
M-m_B-m_{B^*}=- 48^{~+41}_{~-108} ~\mathrm{MeV} ~. 
\end{equation}
 The central value corresponds to  fit of the potential (\ref{V}) in the range $r/a=[2,4]$ that renders the parameters $F=1.3$, $A=0.262(38)$, $d=2.51(27)$. The significant uncertainty of the binding energy captures the statistical errors  
 as well as  various choices for  parametrizing the potential in Fig.  \ref{fig:masses}. The binding energy (\ref{mass}) is consistent with the binding from the previous lattice study \cite{Peters:2016wjm}. 

 The significantly attractive $B\bar B^*$ potential (Fig. \ref{fig:V}) and the resulting  bound state in Fig.  \ref{fig:masses}   could be related to the existence of $Z_b$ in experiment. The   reliable relation between both will be possible only when simplifications employed here will be overcome in the future simulations.  The $Z_b(10610)$ was found as a virtual bound state  slightly below threshold  by the re-analysis of the experimental data  \cite{Wang:2018jlv} when the coupling to bottomonium light-meson channels was turned off \cite{Wang:2018jlv}.  
 
 \section{Conclusions }    

We find a sizable attractive interaction between $B$ and $\bar B^*$ in the $Z_b^+$ channel for separations $r=[0.1,0.4]~$ fm and  this is the most robust conclusion of our lattice QCD simulation. Using further severe simplifying assumptions, we find a bound state   below $B\bar B^*$ threshold, which could be related to $Z_b$ resonances by Belle.  The   reliable relation between both will be possible only when simplifications employed here will be overcome in the future simulations.

\vspace{0.3cm}

\textbf{Acknowledgments} 

\vspace{0.1cm}

We thank   G. Bali,   V. Baru, P. Bicudo,  N. Brambilla, E. Braaten, M. Karliner,  R. Mizuk, A. Peters and  M. Wagner for valuable discussions.    S.P. acknowledges support by Research Agency ARRS (research core funding No. P1-0035 and No. J1-8137) and DFG grant No. SFB/TRR 55. H.B. acknowledges support from the Scientific and Technological Research Council of Turkey (TUBITAK) BIDEB-2219 Postdoctoral Research Programme.


\end{document}